\documentclass[sigconf]{acmart}

\usepackage{enumitem}

\copyrightyear{2023}
\acmYear{2023}
\setcopyright{acmlicensed}\acmConference[SIGGRAPH '23 Conference
Proceedings]{Special Interest Group on Computer Graphics and
Interactive Techniques Conference Conference Proceedings}{August
6--10, 2023}{Los Angeles, CA, USA}
\acmBooktitle{Special Interest Group on Computer Graphics and
Interactive Techniques Conference Conference Proceedings (SIGGRAPH '23
Conference Proceedings), August 6--10, 2023, Los Angeles, CA, USA}
\acmPrice{15.00}
\acmDOI{10.1145/3588432.3591508}
\acmISBN{979-8-4007-0159-7/23/08}

\usepackage{color}
\usepackage{xcolor}
\usepackage{graphicx}
\usepackage{amsmath, amsthm, amsfonts, amssymb}
\usepackage{textcomp}
\usepackage{epstopdf}
\usepackage{multirow}
\usepackage{wrapfig}
\usepackage{subfig}
\usepackage[ruled]{algorithm2e} 
\usepackage{cleveref}
\usepackage{bm}

\SetAlFnt{\small}
\SetAlCapFnt{\small}
\SetAlCapNameFnt{\small}
\SetAlCapHSkip{0pt}

\newcommand{\eg}{e.g.}

\definecolor{green}{rgb}{0, 0.5, 0}
\definecolor{orange}{rgb}{0.6, 0.3, 0.1}
\definecolor{red}{rgb}{1.0, 0.0, 0.0}
\definecolor{teal}{rgb}{0.0, 0.4, 0.4}
\definecolor{purple}{rgb}{0.65,0,0.65}
\definecolor{saffron}{rgb}{0.95,0.75,0.2}
\definecolor{turquoise}{rgb}{0.0,0.5,0.5}
\definecolor{brown}{rgb}{0.5, 0.16, 0.16}
\definecolor{brickred}{rgb}{.6, .2 .1}
\definecolor{coral}{rgb}{1,0.45,0.33}
\definecolor{newcolor}{rgb}{.8,.349,.1}

\begin{document}

\renewcommand\shortauthors{}

\title{UrbanBIS: a Large-scale Benchmark for Fine-grained Urban Building Instance Segmentation}

\author{Guoqing Yang}
\email{yanggq2020@gmail.com}
\affiliation{
\institution{Shenzhen University}
\country{China}
}
\author{Fuyou Xue}
\email{fullcyxuc@gmail.com}
\affiliation{
\institution{Shenzhen University}
\country{China}
}
\author{Qi Zhang}
\email{qi.zhang.opt@gmail.com}
\affiliation{
 \institution{Shenzhen University}
 \country{China}
}
\author{Ke Xie}
\email{ke.xie.siat@gmail.com}
\affiliation{
 \institution{Shenzhen University}
 \country{China}
}
\author{Chi-Wing Fu}
\email{philip.chiwing.fu@gmail.com}
\affiliation{
 \institution{The Chinese University of Hong Kong}
 \country{China}
}
\author{Hui Huang}
\email{hhzhiyan@gmail.com}
\authornote{Corresponding author: Hui Huang (hhzhiyan@gmail.com)}
\affiliation{
 \institution{Shenzhen University}
 \country{China}
}
\renewcommand\shortauthors{Guoqing Yang, Fuyou Xue, Qi Zhang, Ke Xie, Chi-Wing Fu, and Hui Huang}
\begin{abstract}

We present the \emph{UrbanBIS} benchmark for large-scale 3D urban understanding, supporting practical urban-level semantic and building-level instance segmentation. UrbanBIS comprises six real urban scenes, with 2.5 billion points, covering a vast area of 10.78 $km^{2}$ and 3,370 buildings, captured by 113,346 views of aerial photogrammetry. Particularly, UrbanBIS provides not only semantic-level annotations on a rich set of urban objects, including buildings, vehicles, vegetation, roads, and bridges, but also instance-level annotations on the buildings. Further, UrbanBIS is the first 3D dataset that introduces fine-grained building sub-categories, considering a wide variety of shapes for different building types. Besides, we propose \emph{B-Seg}, a building instance segmentation method to establish UrbanBIS. B-Seg adopts an end-to-end framework with a simple yet effective strategy for handling large-scale point clouds. Compared with mainstream methods, B-Seg achieves better accuracy with faster inference speed on UrbanBIS. In addition to the carefully-annotated point clouds, UrbanBIS provides high-resolution aerial-acquisition photos and high-quality large-scale 3D reconstruction models, which shall facilitate a wide range of studies such as multi-view stereo, urban LOD generation, aerial path planning, autonomous navigation, road network extraction, and so on, thus serving as an important platform for many intelligent city applications. UrbanBIS and related code can be downloaded at https://vcc.tech/UrbanBIS.
\end{abstract}



\begin{teaserfigure}
	\centering
	\includegraphics[width=\linewidth]{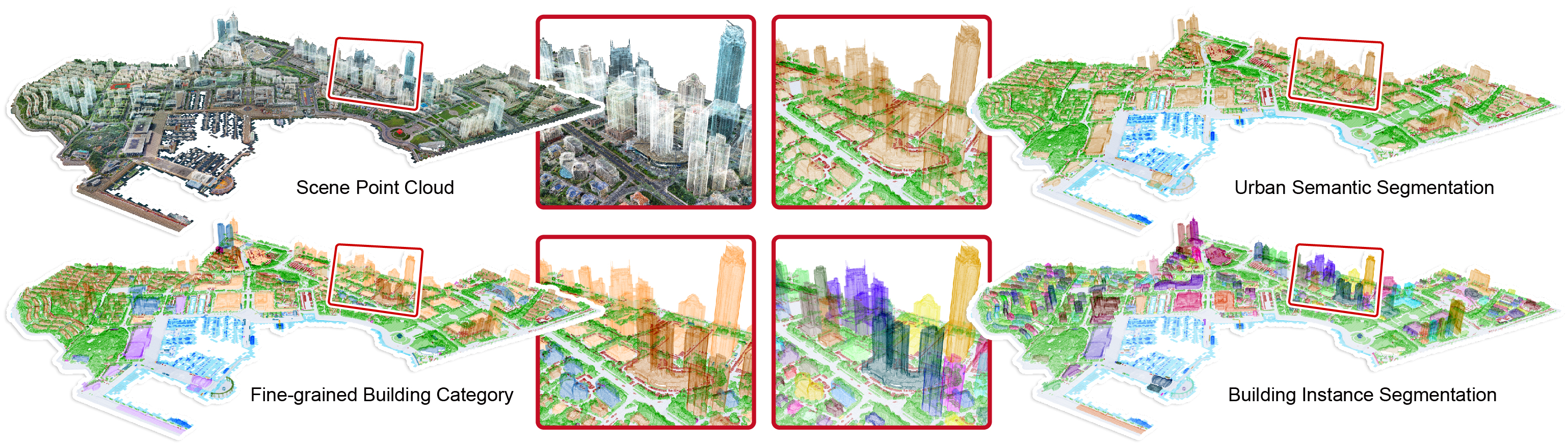}
 \caption{UrbanBIS provides 2.5 billion 3D point samples over six scenes, covering a vast area of 10.78 km$^2$.  Particularly, this large 3D dataset is annotated not only in the urban semantic level (top right) but also in the building instance level (bottom right) with fine-grained building categories (bottom left).
 }
	\label{fig:teaser}
\end{teaserfigure}

\maketitle

\section{Introduction}\label{sec:intro}
\begin{figure*}
  \centering
  \includegraphics[width=0.98\linewidth]{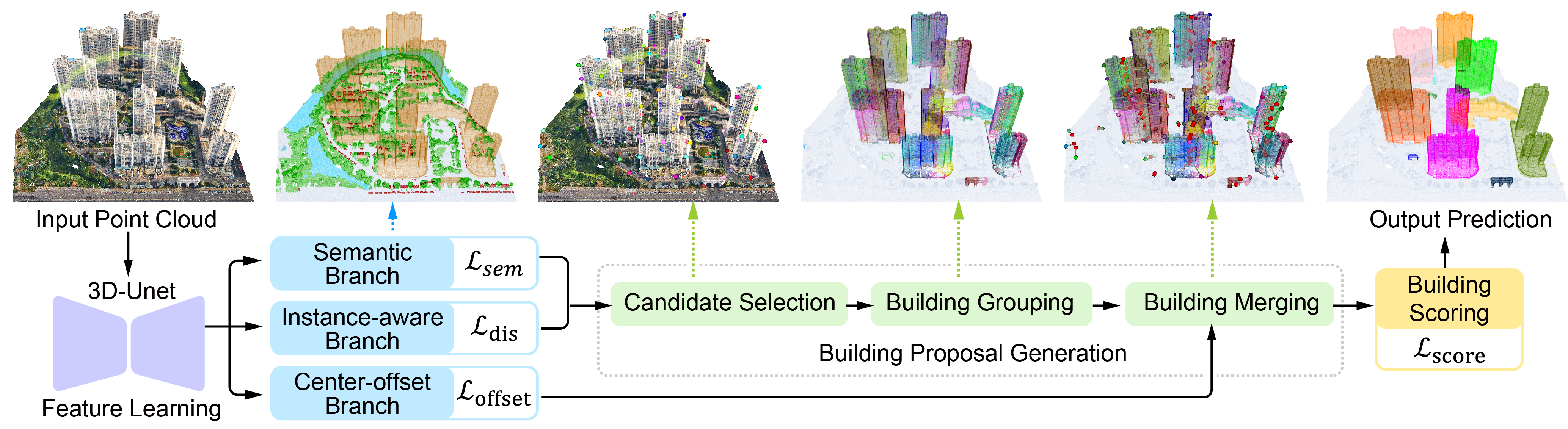}
  \vspace{-3mm}
  \caption{The pipeline of B-Seg:
(i) backbone network and feature learning for point-wise feature extraction and sub-task predictions;
(ii) building proposal generation to produce candidate building instances; and 
(iii) building scoring to filter the candidate instances and produce the final building instances.}
  \label{fig:pipeline}
 \vspace{-4mm}
\end{figure*}

3D machine learning is an emerging research topic, drawing great attention in recent years, as it facilitates a wide range of downstream tasks and applications.
However, existing works focus mainly on tasks at the object level (e.g., object recognition, parts segmentation, shape synthesis, etc.)~\cite{uy2019scanobject,mo2019partnet}, at the indoor scene level (e.g., semantic segmentation,
instance segmentation, floor plan recognition, etc.)~\cite{roberts2021hypersim}, and in outdoor road level (e.g., object detection, autonomous driving, etc.)~\cite{kitti, aygun2021panoptic, zhou2020joint3d, golovinsk2009shape} or mainly focuses on buildings~\cite{selvaraju2021buildingnet}.
The bottleneck comes mainly from the need for datasets and the data annotations, required to train the machine learning models.
In this work, we focus on building an urban-level dataset to facilitate 3D machine learning at the urban level, aiming to support large-scale urban-level 
analysis and understanding~\cite{lafarge2012creating}.

So far, urban datasets are acquired mainly by LiDAR~\cite{behley2019semantickitti,boyko2014cheaper} or UAV photography~\cite{li2020campus3d, gao2021sum}.
The data collection and annotation processes involve tremendous costs, so only few real-world urban-level datasets have been released,~\eg, SensatUrban~\cite{hu2022sensaturban} and STPLS3D~\cite{chen2022stpls3d}. Especially, the recent urban 3D point-cloud datasets~\cite{hu2022sensaturban, chen2022stpls3d} provide only one to three scenes over relatively small areas.
Also, most existing datasets provide only semantic-level annotations without instance-level annotations for real large-scale urban scenes, only small-scale dataset InstanceBuilding~\cite{chen2022mvs} and some datasets containing virtual scenes provide instance information~\cite{chen2022stpls3d, griffiths2019synthcity}.
Furthermore, they provide only urban-level semantics without fine-grained semantic information, e.g., building categories, thereby limiting the applications of the datasets.

Beyond the prior datasets, we propose UrbanBIS (Fig.~\ref{fig:teaser}), the \emph{largest real 3D urban dataset} that we are aware of.
Compared with the latest urban datasets SensatUrban~\cite{hu2022sensaturban} and STPLS3D~\cite{chen2022stpls3d}, UrbanBIS provides \emph{2.5 billions of annotated 3D point samples} and \emph{six real urban scenes}, covering a total area of \emph{$10.78km^2$} in different cities; see Section \ref{sec:comparison} for a quantitative comparison between UrbanBIS and existing datasets.
Very importantly, UrbanBIS provides \emph{both} semantic-level and instance-level annotations, as well as \emph{fine-grained building categories}, which would facilitate many high-level tasks and applications on 3D machine learning.

Building UrbanBIS is a very tedious and expensive process,
involving UAV image acquisition (0.5TB of 113,346 photos), 3D urban reconstruction, 3D mesh annotations, point cloud sampling (2.5 billion points in 3D), and fine-grained building annotations (3,370 buildings), consuming about 1,600 man-hours of annotation works in total.
Besides, we propose a new instance segmentation method called B-Seg for handling large-scale point-cloud scenes (see Fig.~\ref{fig:pipeline}).
Compared with the existing 3D instance segmentation methods,~\eg,~\cite{jiang2020pointgroup,chen2021hierarchical,vu2022softgroup}, which typically adopt a time-costly point-wise clustering, we formulate an efficient pipeline in B-Seg to exploit point feature similarity and 
segment the points through a relation matrix, thus avoiding the processing of all points in the whole large-scale urban scenes.
Compared with existing methods, B-Seg achieves better performance and higher inference speed on the large-scale UrbanBIS dataset.

Overall, the contribution of our work is three-fold: (i) UrbanBIS, a large-scale real-world 3D urban dataset annotated with fine-grained building categories and instance segmentation information; (ii) B-seg, a fast and accurate new 3D point cloud instance segmentation method for buildings in large-scale urban scenes and (iii) A 3D urban platform with a rich variety of data that has great potential for developing many applications such as 3D reconstruction, depth prediction, multi-view stereo, and aerial path planning.
\section{Related Work}\label{sec:rw}

\subsection{Indoor datasets for segmentation}
Several datasets~\cite{armeni2017s3dis, dai2017scannet, xiao2013sun3d, roberts2021hypersim, rozenberszki2022scannet200} have been built for indoor scene parsing and understanding. Yet, they are mainly for object classification~\cite{uy2019scanobject}, part segmentation~\cite{mo2019partnet}, and indoor-scene semantic segmentation~\cite{roberts2021hypersim}.

Due to the expensive cost and difficulties of preparing the annotations, only a few indoor datasets~\cite{armeni2017s3dis, dai2017scannet} provide instance-level masks for supporting the instance segmentation task. Dai et al.~\shortcite{dai2017scannet} prepared ScanNet, an RGB-D video dataset, containing 2.5M views in 1,513 scenes annotated with 3D camera poses for surface reconstructions and segmentations. This dataset was obtained using commercial cameras, consisting of 21 categories and facilitating both 2D and 3D semantic and instance segmentation.
Another common large-scale indoor dataset is S3DIS~\cite{armeni2017s3dis}, which covers a total area of 6000$\textit{m}^{2}$, which was acquired using the Matterport cameras.
Beyond the 20 semantic categories in ScanNet~\cite{dai2017scannet}, Rozenberszki et al.~\shortcite{rozenberszki2022scannet200} recently proposed ScanNet200, a new dataset with 200 semantic categories for fine-grained 3D indoor scene understanding.

\subsection{Urban datasets}
Urban scenes have unique features such as large-scale and complex lighting conditions, for which LiDAR and multi-view stereo are the major technologies for acquiring urban-level scene data.

\textbf{LiDAR}. According to the acquisition platform that carries the LiDAR~\cite{behley2019semantickitti}, the scanning can be roughly divided into mobile laser scanning (MLS)~\cite{okland, xavier2018PL3D, tan2020toronto3d}, terrestrial laser scanning (TLS)~\cite{timo2017semantic3d}, and airborne laser scanning (ALS)~\cite{varney2020dales, michael2021h3d}. 
Okland~\cite{okland} is a point cloud dataset collected by MLS, consisting of 44 semantic categories, in which 24 of the categories contain less than 1,000 point clouds.
Semantic3D~\cite{timo2017semantic3d} is collected via TSL. It comprises eight semantic categories and provides two different benchmarks on 2D images and 3D points.
The H3D dataset~\cite{michael2021h3d} adopts a multimodal data acquisition method with a LiDAR and an optical camera installed on an airborne platform. It consists of 11 semantic categories. While LiDAR-based scanning gives promising results, its drawbacks are obvious.
First, the scanning angle is fixed and there is an occlusion issue. Second, the scanning density is inversely proportional to the object-sensor distance, leading to varying point patterns and sparse points for distant objects in the results.

\textbf{Multi-view Stereo}.
Recently, researchers started to collect 3D urban data using multi-view stereo~\cite{li2020campus3d, gao2021sum}. 
Swiss3DCity~\cite{gulcan2021swiss} uses an array of high-resolution cameras to capture images and provides more complete and dense point clouds for three Swiss cities using UAV photogrammetry.
It also provides point clouds of different resolutions for research purposes.
The SensatUrban dataset~\cite{hu2022sensaturban} was proposed to mitigate the small-scale and limited semantic annotations in the current dataset, providing three billion points that cover an area of 7.6$km^{2}$ in three different urban scenes.
It provides annotations of 13 semantic categories. 
The STPLS3D dataset~\cite{chen2022stpls3d} contains instance-labeled urban virtual scenes and real scenes labeled only with semantic information.
While the aforementioned datasets have contributed greatly to the development of research on urban understanding, they are annotated mainly with semantic information and lack instance- or building-level annotations (except STPLS3D and SynthCity~\cite{griffiths2019synthcity}, which contain instance labels only for virtual scenes). InstanceBuilding~\cite{chen2022mvs} is a real urban scene dataset with building instances, but it is small in scale and has only two semantic categories.

In addition to the above urban scene datasets aimed for segmentation or detection, some urban scene datasets were proposed for other domains.
Lin et al.~\shortcite{lin2022UrbanScene3D} built UrbanScene3D, a large-scale urban scene dataset for path planning and 3D reconstruction, providing ten virtual scenes and six real scenes.
Besides, there are 3DCityDB~\cite{yao20183dcitydb} and the large-scale 3D geospatial data~\cite{biljecki2019geodata} proposed for urban management and urban planning, providing visualizations and management of urban scenes.

\subsection{3D Instance segmentation methods}
Existing 3D instance segmentation methods are oriented mainly to indoor scenes.
They can be roughly divided into proposal-based methods and proposal-free methods.
Proposal-based methods are kind of a top-down approach that first generates region proposals such as bounding boxes then predicts instance masks~\cite{GSPN, hou20193dsis}. 
3D-BotNet~\cite{3D-BoNet} directly regresses 3D bounding boxes for object candidates with a multi-criteria loss. 
3D-MPA~\cite{3D-MPA} samples proposals from shifted centroids and uses a graph neural network~\cite{DGCNN} to enhance the features.
GICN~\cite{liu2020gicn} regards centroids of instances as a Gaussian distribution.
Overall, these methods show results with good objectness, yet they lack efficiency.
Also, the quality of their results highly depends on the proposals.

Proposal-free methods use a bottom-up style that directly extracts point features then clusters points into object instances~\cite{ASIS,JSNet,JSIS3D,OccuSeg,liang2021treenet,chen2021hierarchical}.
SGPN~\cite{SGPN} introduces a feature matrix to represent point-pair similarity to aid the clustering.
MTML~\cite{MTML} groups instances from discriminative feature embedding with mean-shift clustering.
PointGroup~\cite{jiang2020pointgroup} clusters object instances by simultaneously considering two different sets of point coordinates.
SoftGroup~\cite{vu2022softgroup} improves PointGroup~\cite{jiang2020pointgroup} with the soft semantic scores. 
Some other methods~\cite{DyCo3D,DKNet} adopt kernel-based strategies to exploit kernel features.

So far, existing 3D instance segmentation methods aim to handle mainly indoor scenes with small household-level objects. It remains unclear whether they can effectively process and segment large and complex urban-level scenes.
Especially, modern buildings have diverse shapes, sizes, and appearances.
\begin{table}
  \centering
  \caption{Statistics of the collected aerial photos in UrbanBIS.}
  \resizebox{\linewidth}{!}{
  \begin{tabular}{c@{\hspace{2mm}}|c@{\hspace{2mm}}c@{\hspace{2mm}}c@{\hspace{2mm}}c}
    \toprule
    Scene & Area ($km^2$) & Image (\#) & Resolution (pixels) & Size (GB)\\
    \midrule
    Qingdao & 2.31 & 98,373 & $6,000\times4,000$ & 189.9\\
    Wuhu & 2.92 & 1,053 & $14,204\times10,652$ & 140\\
    Longhua & 2.41 & 11,307 & $8,192\times5,640$ & 162\\
    Yuehai & 1.60 & 1,030 & $5,472\times3,648$ & 8.33\\
    Lihu & 1.46 & 728 & $6,000\times4,000$ & 6.9\\
    Yingrenshi & 0.08 & 855 & $5,472\times3,648$ & 6.97\\
    \bottomrule
    Total & 10.78 & 113,346 & - & 514.1\\
    \bottomrule
  \end{tabular}
  }
  \label{tab:basic}
  \vspace{-2.5mm}
\end{table}

\section{The UrbanBIS dataset}\label{sec:overview}

\subsection{UrbanBIS acquisition and annotations}
\textbf{Data acquisition.} \
To obtain complete scene data for a large-scale urban area, we adopt aerial photogrammetry to collect images due to the flexibility of the UAV platform. To speed up the reconstruction process and to improve the quality of the reconstructed models, we adopt~\cite{zhou2020offsite} to generate trajectories instead of using conventional oblique photogrammetry. The acquisition devices in our setting include the \textit{DJI PHANTOM} 4 \textit{RTK} drone\footnote{\url{https://www.dji.com/phantom-4-rtk}} with the built-in camera and the \textit{DJI M}300\textit{RTK} drone\footnote{\url{https://www.dji.com/matrice-300}} loaded with five \textit{HD PSDK} 102S aerial cameras.
Table~\ref{tab:basic} shows the basic information of the aerial photos (0.5 TB) we collected in six different scenes.

\begin{figure*}[t]
	\centering
	\includegraphics[width=\linewidth]{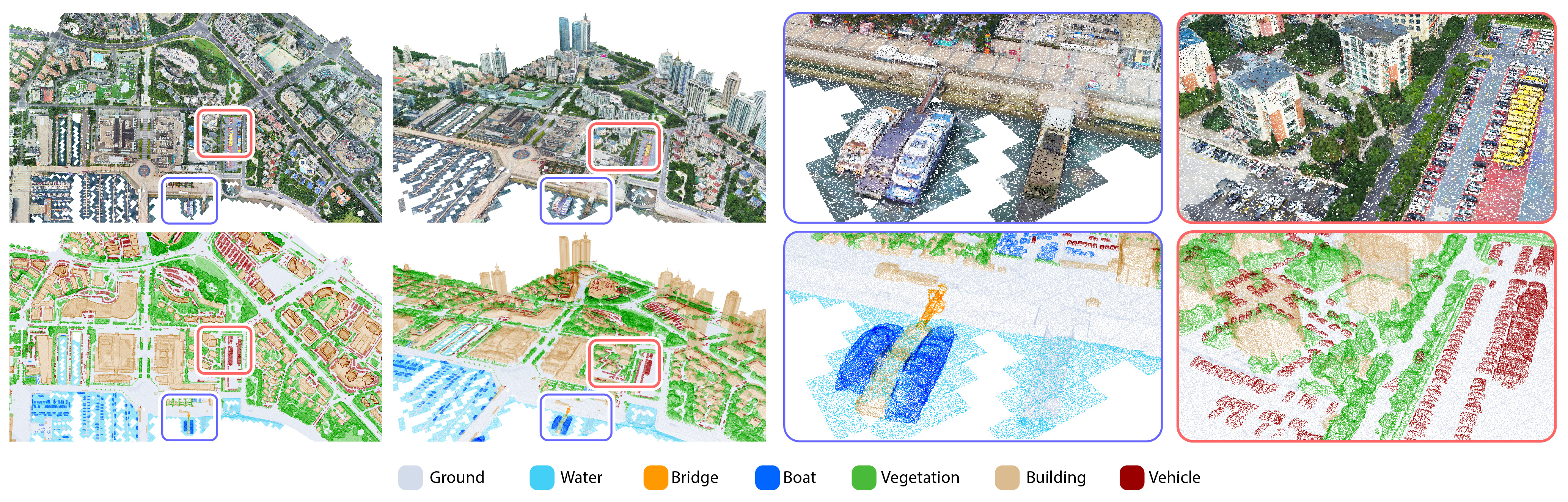}
	\caption{The examples of the semantic labeling of different blocks in UrbanBIS. The color legend is shown at the bottom. See more in the Supp.}
	\label{fig:semantic}
 \vspace{-2mm}
 \end{figure*}

\textbf{Mesh annotation.} 
We follow conventional datasets ~\cite{gao2021sum} to define the semantic categories.
Overall, UrbanBIS provides two kinds of semantic information: (i) urban-level semantics and (ii) building-level semantics.
Below, we list the urban-level semantic categories:
(i) \emph{Ground}: Impervious surface, roads, parking, etc.;
(ii) \emph{Water}: Lake, river, sea, etc.;
(iii) \emph{Boat}: Cruise ships, small boats, etc.;
(iv) \emph{Vegetation}: Trees, lawn, bushes, etc.;
(v) \emph{Bridge}: Bridges across rivers or in the parks;
(vi) \emph{Vehicle}: Cars, buses, bikes, etc.; and
(vii) \emph{Building}: The instantiated categories, including various man-made buildings.
Note that some datasets put grass into the ground category while others put it into vegetation.
Here, we regard lawns and bushes as vegetation and regard wild grasses on the ground as ground. 
Fig.~\ref{fig:semantic} shows the semantic definition in UrbanBIS. The detailed number of mesh faces for each urban-semantic category can be found in the supplementary material.

\textbf{Point cloud sampling.}
As the reconstructed 3D meshes lack water tightness, performing machine learning directly on them may lead to serious deviations in results.
Hence, we generate point samples on the reconstructed meshes to represent the 3D scene. In detail, we adopt \textit{CloudCompare}\footnote{\url{https://www.cloudcompare.org/}} to generate point samples in each 3D scene. Considering that the sampled points should be consistent with the geometry of the reconstructed 3D mesh and also uniform in distribution, we choose to sample the points based on the surface density of the mesh. Here, we set the number of sample points per square meter to 80 and represent each point by a 6D feature with 3D coordinates and RGB information.
Table~\ref{tab:point cloud statics} reports the number of points for different urban semantics in each scene.

\begin{table}
  \centering
  \caption{UrbanBIS provides 2.5 billion sampled points in six scenes.}
  \vspace{-0.3cm}
  \resizebox{0.9\linewidth}{!}{
  \begin{tabular}{c@{\hspace{2mm}} c@{\hspace{2mm}} c@{\hspace{2mm}} c@{\hspace{2mm}} c@{\hspace{2mm}} c@{\hspace{0mm}} c}
    \toprule
    Category & Qingdao & Wuhu & Longhua & Yuehai & Lihu & Yingrenshi\\
    \midrule
  Size (GB) & 26.5 & 27.8 & 29.1 & 17.5 & 11.5 & 0.92\\
  Building (\#) & 269.59M & 285.28M & 256.39M & 117.98M & 65.18M & 14.97M\\
  Ground (\#) & 114.22M & 133.32M & 158.62M & 69.60M & 80.54M & 4.39M\\
  Water (\#) & 11.46M & 20.95M & 0.26M & 3.86M & 2.46M & 0\\
  Boat (\#) & 4.20M & 409 & 852 & 0 & 2,490 & 0\\
  Vegetation (\#) & 179.50M & 175.69M & 225.50M & 197.83M & 104.09M & 1.66M\\
  Vehicle (\#) & 15.05M & 8.24M & 11.35M & 1.16M & 2.08M & 0.85M\\
  Bridge (\#) & 37,074 & 1.61M & 1.77M & 2.93M & 0.78M & 0.35M\\
    \bottomrule
    Total (\#) & 594.06M & 625.08M & 653.90M & 393.37M & 255.12M & 22.22M\\
    \bottomrule
  \end{tabular}}
  \label{tab:point cloud statics}
  \vspace{-0.3cm}
\end{table}

\subsection{Fine-grained Building-level information in UrbanBIS}
\label{sec:building_details}

\begin{figure*}[t]
	\centering
	\includegraphics[width=\linewidth]{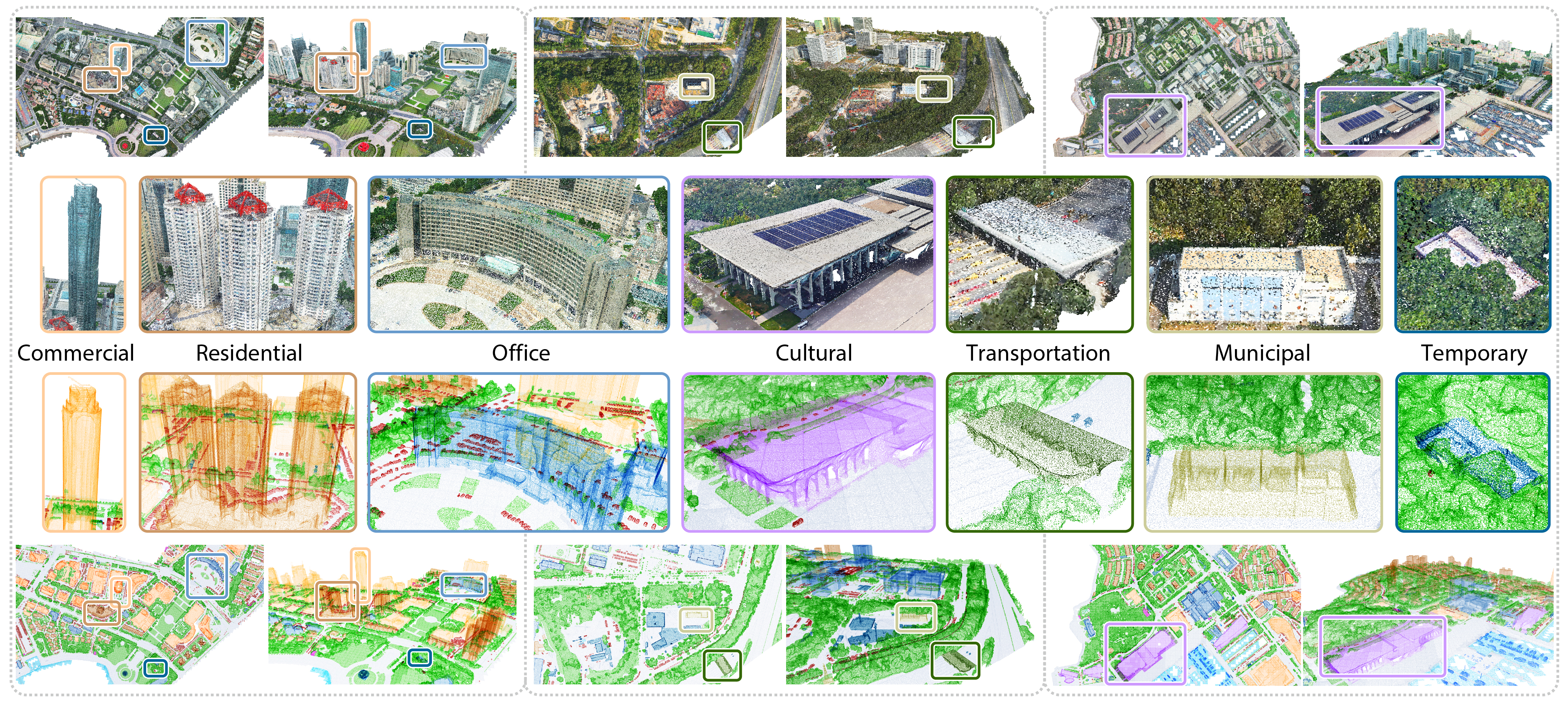}
	\caption{
UrbanBIS provides fine-grained building-level information, including segmentation information on building instances and semantics information about building categories (from left to right):
Commercial, Residential, Office, Cultural, Transportation, Municipal and Temporary.
	}
	\label{fig:diversity}
 \vspace{-0.3cm}
\end{figure*}

Concerning the building-level semantics, UrbanBIS provides not only segmentation information on building instances but also semantic information on buildings.
Buildings can be divided into different categories based on their functions.
According to the scheme proposed by the Department of Engineering Quality and Safety Supervision, Ministry of Housing and Urban-Rural Development~\shortcite{national}, buildings can be divided into 28 categories,~\eg, residential, commercial, cultural, etc.
Buildings of different functions usually have large variations in appearance and shape (see Fig.~\ref{fig:diversity}).
To provide a more detailed description of urban scenes, we further classify buildings into fine-grained sub-categories. By merging the building types based on the building functions, we consider the following 7 building categories:
(i) \emph{Commercial}: including commercial, catering, entertainment, hotel, financial, and exhibition buildings;
(ii) \emph{Residential}: including residential and dormitory buildings;
(iii) \emph{Office}: including office, research, medical, hygienic, and telecommunication buildings;
(iv) \emph{Cultural}: including cultural, museums, gyms, and religious buildings;
(v) \emph{Transportation}: including subway entrances and bus stops;
(vi) \emph{Municipal}: including waste disposal stations and electrical facilities; and
(vii) \emph{Temporary}: including temporary and garden buildings.
Also, we consider another aspect of classifying buildings: \emph{low-rise\/} (under 24m); \emph{high-rise\/} (24m to 100m); and \emph{super high-rise\/} (over 100m).
Please refer to Table~\ref{tab:building_type_height_distribution} for the detailed number of buildings in UrbanBIS for each category.

\begin{table}
	\caption{The distribution of building categories (middle) and building heights (right) in each scene in the UrbanBIS dataset. Co, Re, Of, Cu, Tr, Mu, and Te stand for Commercial, Residential, Office, Cultural, Transportation, Municipal, and Temporary, respectively.  L, H, and SH stand for Low-rise, High-rise, and Super High-rise, respectively. We ignore buildings that are difficult to determine their functions during the data category annotation.}
 \vspace{-0.2cm}
    \resizebox{1\linewidth}{!}{
	\begin{tabular}{c|ccccccc|ccc}
		\toprule
		Scene & Co & Re & Of & Cu & Tr & Mu & Te & L & H & SH\\
		\midrule
		Qingdao & 64 & 238 & 26 & 8 & 18 & 106 & 124  & 554 & 77 & 27\\
		Wuhu & 24 & 813 & 32 & 7 & 0 & 47 & 117  & 1000 & 73 & 28\\
        Yuehai & 7 & 55 & 39 & 16 & 1 & 12 & 114  & 220 & 35 & 1\\
        Lihu & 1 & 14 & 26 & 7 & 4 & 44 & 211  & 300 & 21 & 1\\
		Longhua & 12 & 274 & 96 & 1 & 17 & 111 & 454  & 844 & 132 & 20\\
		Yingrenshi & 3 & 11 & 10 & 0 & 0 & 4 & 6  & 19 & 18 & 0\\
        \bottomrule
        Total& 111& 1405&229&39&40&324&1026&2937&356&77\\
        \bottomrule
	\end{tabular}
    }
	\label{tab:building_type_height_distribution}
 \vspace{-0.2cm}
\end{table}
 
\begin{table*}
  \centering
  \caption{Comparing existing 3D urban segmentation datasets.
  MLS, TLS, and ALS stand for mobile laser scanning, terrestrial laser scanning, and aerial laser scanning, respectively.
  In the Application column, SS, IR, SC, and IS stand for semantic segmentation, image reconstruction, scene completion, and instance segmentation, respectively.
  Ptgy is Photogrammetry and sub-cat. is Sub-category.
  Only real scenes are considered in this table.
  }
  \vspace{-0.3cm}
  \resizebox{\textwidth}{!}{
  \begin{tabular}{c@{\hspace{2mm}}c@{\hspace{2mm}}c@{\hspace{2mm}}c@{\hspace{2mm}}c@{\hspace{2mm}}c@{\hspace{2mm}}c@{\hspace{2mm}}c@{\hspace{2mm}}c@{\hspace{2mm}}c}
    \toprule
    Dataset & Year & Acquisition & Data-type & Area/Length & Scene & Points/Triangles & Classes & Application & Annotation\\
    \midrule
    Okland~\cite{okland} & 2009 & MLS & PC & 1.5$km$ & 1 & 1.6 M & 5 & SS & Semantic\\
    Semantic3D~\cite{timo2017semantic3d} & 2017 & TLS & PC & - & 3 & 4000M & 8 & SS & Semantic\\
    Paris-Lille-3D~\cite{xavier2018PL3D} & 2018 & MLS & PC & 1.94$km$ & 2 & 143M & 9 & SS & Semantic\\
    DublinCity~\cite{zol2019dublincity} & 2019 & ALS & PC/Image & 2$km^2$ & 1 & 260M & 13 & SS/IR & Semantic\\
    SemanticKITTI~\cite{behley2019semantickitti} & 2019 & MLS & PC & 39.2$km$ & 1 & 4549M & 25 & SS/SC & Semantic\\
    Toronto-3D~\cite{tan2020toronto3d} & 2020 & MLS & PC & 1.0$km$ & 1 & 78.3M & 8 & SS & Semantic\\
    DALES~\cite{varney2020dales} & 2020 & ALS & PC & 10$km^2$ & 1 & 505.3M & 8 & SS & Semantic\\
    Campus3D~\cite{li2020campus3d} & 2020 & UAV Ptgy & PC/Image & 1.58$km^2$ & 1 & 937.1M & 14 & SS/IS & Hierarchical\\
    Hessigheim 3D~\cite{michael2021h3d} & 2021 & UAV LiDAR/Camera & PC/Mesh & 0.19$km^2$ &1& 125.7M/36.76M & 11 & SS & Semantic\\
    SUM~\cite{gao2021sum} & 2021 & Airplane Camera & Mesh & 4$km^2$ & 1&19M & 6 & SS & Semantic\\
    Swiss3DCities~\cite{gulcan2021swiss} & 2021 & UAV Ptgy & PC & 2.7$km^2$ &3& 226M & 5 & SS & Semantic\\
    SensatUrban~\cite{hu2022sensaturban} & 2022 & UAV Ptgy & PC & 7.46$km^2$ &1& 2847M & 13 & SS & Semantic\\
    STPLS3D~\cite{chen2022stpls3d} & 2022 & UAV Ptgy & PC/Mesh & 1.27$km^2$&1 & 150.4M & 6  & SS & Semantic\\
    InstanceBuilding~\cite{chen2022mvs} & 2022 & UAV Ptgy & Mesh/Image & 0.434$km^2$ & 1 & 7.46M & 2 & IS & Instance\\
    UrbanBIS (Ours) & 2023 & UAV Ptgy & PC/Mesh/Image & 10.78$km^2$ & 5 & 2523.8M/284.3M & 7/8 (Sub-cat.) & SS/IS/IR/SC & Semantic/Instance\\
    \bottomrule
  \end{tabular}}
  \label{tab:comparision}
  \vspace{-0.1cm}
\end{table*}

\subsection{Analysis and comparison}\label{sec:comparison}

As seen from Table~\ref{tab:building_type_height_distribution} (middle) , UrbanBIS contains a rich variety of different building categories.
Also, the six scenes exhibit different compositions of building categories.
Interestingly, both Yuehai and Lihu contain fewer commercial or transportation buildings, but have a lot of office, cultural, and residential buildings, as they are both campuses.
Besides, we can observe that the two campus scenes have a large proportion of temporary buildings, since both scenes contain a large number of prefab houses that are under construction.
The Wuhu scene contains more residential buildings, while Qingdao has more commercial and transportation buildings, showing that Qingdao, as a larger city, is at a higher city development level. 
Current urban datasets often ignore differences across scenes.
This may lead to models trained on specific scenes being difficult to generalize. 
UrbanBIS includes a wider range of scenes and we analyze the similarities between scenes based on the number of buildings in different semantic categories. Table~\ref{tab:correlation} reports the correlation coefficients between scenes, revealing significant differences in their semantic categories, except for Yuehai and Lihu, which are both campus scenes. Researchers can choose to employ scenes that are more similar to the intended use in their studies.

\begin{figure}
	\centering
	\includegraphics[width=0.42\textwidth]{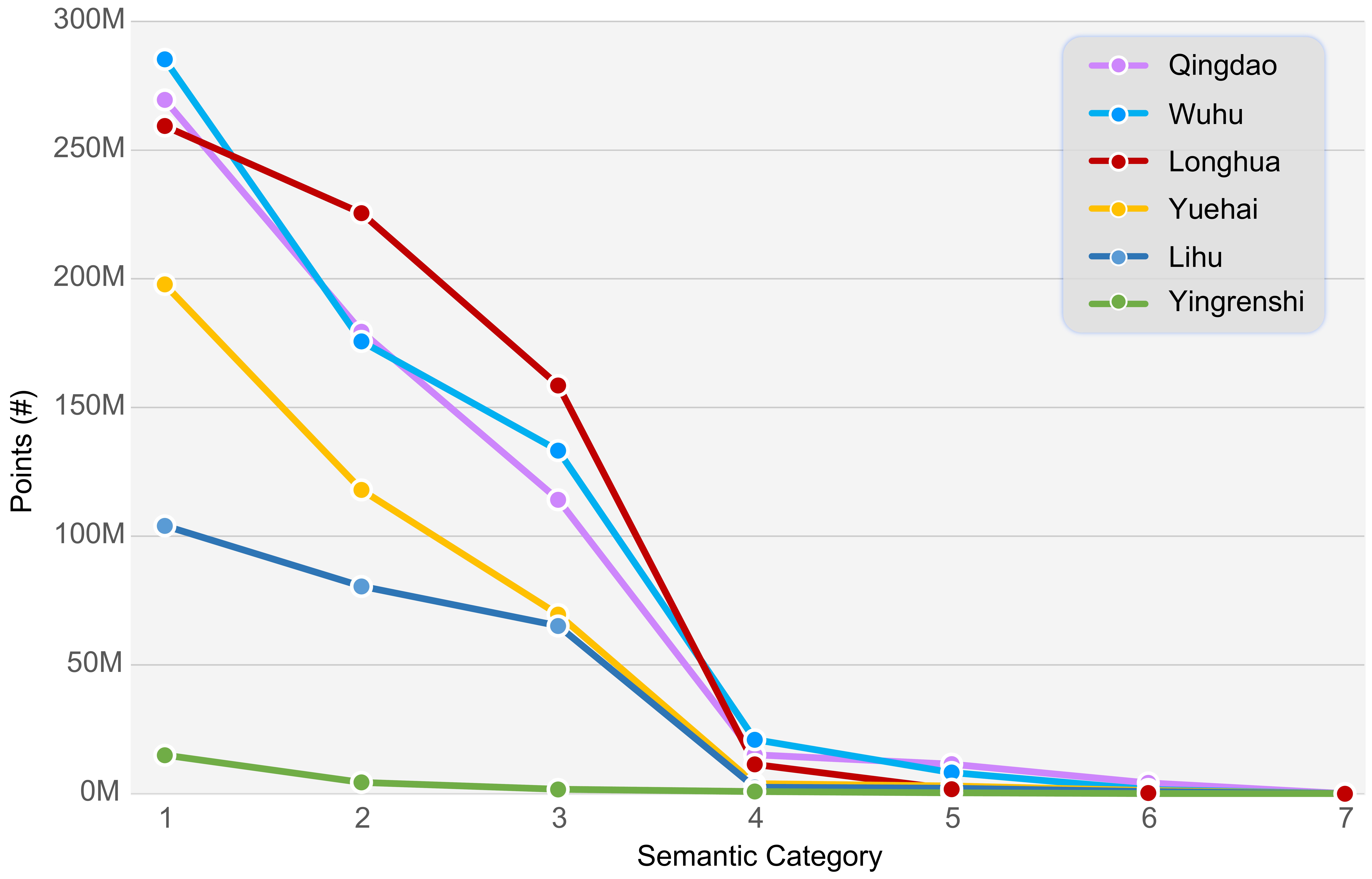}	
 \vspace{-0.3cm}
 \caption{The long tail statistics of different scenes in UrbanBIS. The X-axis represents the 7 semantic categories defined in UrbanBIS in numeric form. They are arranged in descending order of the number of point clouds in each scene separately. }
	\label{fig:longtail}
 \vspace{-0.4cm}
\end{figure}

UrbanBIS is a real-world dataset, providing various semantic categories. Therefore, when designing segmentation models based on this data, the `long-tail' problem that the data may bring needs to be considered. Fig.~\ref{fig:longtail} shows the number of point clouds for each semantic category in UrbanBIS, indicating that the three larger categories account for the vast majority, and current methods still face certain problems in recognizing smaller categories.

\begin{table}[!t]
\centering
  \caption{The correlations among all the scenes in UrbanBIS.}
  \vspace{-0.2cm}
  \resizebox{1\linewidth}{!}{
   \begin{tabular}{c|cccccc}
   \toprule
      & Qingdao & Wuhu  & Longhua & Yuehai & Lihu  & Yingrenshi \\ \midrule
    Qingdao    & 1       & 0.89  & 0.68    & 0.50   & 0.26  & 0.65       \\
    Wuhu       & 0.89    & 1     & 0.47    & 0.34   & -0.05 & 0.66       \\
    Longhua    & 0.68    & 0.47  & 1       & 0.96   & 0.85  & 0.56       \\
    Yuehai     & 0.50    & 0.34  & 0.96    & 1      & 0.88  & 0.53       \\
    Lihu       & 0.26    & -0.05 & 0.85    & 0.88   & 1     & 0.18       \\
    Yingrenshi & 0.65    & 0.66  & 0.56    & 0.53   & 0.18  & 1          \\ \bottomrule
  \end{tabular}}
\label{tab:correlation}
\vspace{-0.3cm}
\end{table}

Compared with existing 3D point cloud datasets, UrbanBIS is the largest real urban dataset that provides both semantic and instance annotations.
As Table~\ref{tab:comparision} shows, UrbanBIS covers an area of 10.78 $km^2$, larger than the latest urban datasets SensatUrban~\cite{hu2022sensaturban} and STPLS3D~\cite{chen2022stpls3d}.
Importantly, UrbanBIS is currently the only large-scale real urban dataset that provides instance annotations and fine-grained building categories.
Besides, it has the richest variety of scenes (six real urban scenes), including three large scenes in different cities (Qingdao, Wuhu, and Longhua), two campus scenes (Yuehai and Lihu), and one small residential area (Yingrenshi).
Hence, UrbanBIS facilitates not only semantic segmentation and instance segmentation but also many high-level tasks and applications,~\eg, image reconstruction and scene completion.

Further, from Table~\ref{tab:building_type_height_distribution} (right), we can see that UrbanBIS contains buildings of various heights.
An intriguing observation is that even though different scenes have different compositions of building categories, they have similar distributions of building heights (if the scene area is sufficiently large), except for Yingrenshi, which is a small scene.
So, we can take Yingrenshi as a test scene in UrbanBIS to explore the domain gap between the train and test samples, as well as the model generalization.
Note also that the supplementary material presents more analysis,~\eg, building density.
See our accompanying video for a vivid understanding of UrbanBIS. 
\section{B-Seg method}
\label{sec:method}
Existing mainstream methods~\cite{jiang2020pointgroup, chen2021hierarchical, liang2021treenet, vu2022softgroup} commonly use clustering algorithms to group foreground points into instance proposals. Our B-Seg is \emph{clustering-free}, enabling higher computational efficiency on processing large-scale urban-level data.
As depicted in Fig.~\ref{fig:pipeline}, B-Seg has three components:
(i) backbone network and feature learning for point-wise feature extraction and sub-task predictions; (ii) building proposal generation to produce and form preliminary building proposals; and (iii) building scoring to evaluate the quality of building proposals and filter out the errors.

\vspace*{2mm}
\noindent
\textbf{(i) Backbone Network and Feature Learning.} \
We use a submanifold sparse convolutional network~\cite{graham20183d} as our backbone due to its strong capability of extracting features from 3D point clouds. 
Then, with the extracted point-wise features, we construct three parallel branches for three different sub-tasks: the semantic branch for semantic segmentation, the center-offset branch for predicting the offset of each point to its associated center, and the instance-aware branch for enhancing the instance segmentation result.
Please refer to the supplementary material for details.

\vspace*{2mm}
\noindent
\textbf{(ii) Building Proposal Generation.} \
Next, we generate instance proposals for buildings with the following three modules:
(i) \emph{Candidate selection}: First, we employ furthest point sampling~\cite{elda1997fps} to randomly select foreground points as candidates for forming building proposals.
In our experiments, we dynamically select one candidate point for every 3,000 foreground points, with a maximum of 100 per block.
(ii) \emph{Building grouping}: We construct a relation matrix, in which the element at $i$-th row, $j$-th column denotes the learned feature distance between the $i$-th foreground point and $j$-th candidate point; the lower the feature distance, the higher the probability that the two points belong to the same building.
We then use the $argmin$ function to obtain the building proposal label for each foreground point.
(iii) \emph{Building Merging}: Since there are still a large number of candidate points, the same building instance may be covered by multiple candidates.
So, we offset each candidate point towards its predicted instance center and then merge candidate points and instance proposals for the same building.

\begin{table*}[t!]
\caption{Benchmark results. Comparing the 3D point cloud instance segmentation performance of our method with state-of-the-art methods 
on the train and test sets of UrbanBIS.
All methods are trained and tested under the same scene.
Campus combines Yuehai and Lihu.
T stands for the time cost for inference.}
\resizebox{0.9\textwidth}{!}{
\begin{tabular}{c|cccc|cccc|cccc|cccc}
\toprule
& \multicolumn{4}{c|}{Qingdao} & \multicolumn{4}{c|}{Wuhu} & \multicolumn{4}{c|}{Longhua} & \multicolumn{4}{c}{Campus}\\ 
\midrule
Method & AP& AP50 & AP25 & T (s) & AP & AP50 & AP25 & T (s) & AP & AP50 & AP25 & T (s) & AP & AP50 & AP25 & T (s)\\ 
\midrule
PointGroup~\cite{jiang2020pointgroup}&0.364&0.512&0.578&9.80&0.502&0.662&0.748&5.90&0.318&0.443&0.556&5.73&0.117&0.235&0.455&3.65\\
HAIS~\cite{chen2021hierarchical}&0.320&0.465&0.506&7.11&0.383&0.616&0.711&3.62&0.159&0.249&0.350&3.17&0.002&0.012&0.146&3.26\\
SoftGroup~\cite{vu2022softgroup}&0.383&0.446&0.487&6.55&0.536&0.649&0.721&3.61&0.151&0.199&0.300&3.06&0.253&0.364&0.439&2.16\\
DyCo3D~\cite{DyCo3D}&0.285&0.376&0.498&5.20&0.470&0.620&0.732&3.04&0.020&0.045&0.196&1.77&0.029&0.063&0.180&1.67\\
DKNet~\cite{DKNet}&0.383&0.434&0.474&2.15&0.474&0.575&0.650&1.20&0.077&0.154&0.253&1.78&0.044&0.109&0.251&0.88\\
B-Seg (Ours) &\textbf{0.453}&\textbf{0.550}&\textbf{0.672}&\textbf{1.19}&\textbf{0.549}&\textbf{0.674}&\textbf{0.767}&\textbf{0.99}&\textbf{0.402}&\textbf{0.513}&\textbf{0.618}&\textbf{1.16}&\textbf{0.261}&\textbf{0.386}&\textbf{0.535}&\textbf{0.74}\\ 
\bottomrule
\end{tabular}}
\label{tab:result1}
\end{table*}

\vspace*{2mm}
\noindent
\textbf{(iii) Building Scoring.} \
The previous module may mistakenly produce some wrong proposals, which may affect the quality of the final instance predictions.
So, we adopt ScoreNet~\cite{jiang2020pointgroup} to further predict a score for each proposal. 
In detail, we pass each proposal into a tiny 3D sparse convolution U-Net with an instance-aware pooling and a softmax layer to predict a score and filter out building proposals with scores less than 0.1.

In the end, we train the B-Seg framework end-to-end from scratch with an overall optimization objective with four loss terms (see Fig.~\ref{fig:pipeline}):
$\mathcal{L}_{sem}$ for semantic segmentation,
$\mathcal{L}_{offset}$ for optimizing the center-offset vector prediction,
$\mathcal{L}_{dis}$ for learning the instance-aware features, and 
$\mathcal{L}_{score}$ for learning the instance-proposal scores.

\section{Experimental Results}\label{sec:results}

\begin{table*}[t!]
	\caption{3D instance segmentation performance of various methods on the fine-grained building sub-categories in UrbanBIS.
Co, Re, Of, Cu, Tr, Mu, Te, L, H, and SH stand for Commercial, Residential, Office, Cultural, Transportation, Municipal, Temporary, Low-rise, High-rise, and Super High-rise, respectively.
  }
\resizebox{0.75\textwidth}{!}{
\begin{tabular}{c|c|ccccccc|ccc}
		\toprule
		Scene & Method& Co & Re & Of & Cu & Tr & Mu & Te & L & H  & SH \\
		\midrule
		\multirow{6}{*}{Qingdao} & PointGroup~\cite{jiang2020pointgroup}   & 0.831& 0.946& 0.798& 0.620& 0.329& 0.527& 0.480& 0.780& 0.867& 0.918\\
		\cline{2-12} & HAIS~\cite{chen2021hierarchical}         & 0.913& 0.981& 0.945& 0.662& 0.621& 0.510& 0.329& 0.872& 0.920& 0.970\\
		\cline{2-12} & SoftGroup~\cite{vu2022softgroup}    & 0.823& 0.94& 0.71& 0.479& 0.645& 0.413& 0.242& 0.423& 0.856& 0.859\\
		\cline{2-12} & DyCo3D~\cite{DyCo3D}       & 0.794& 0.909& 0.691& 0.502& 0.224& 0.377& 0.305& 0.699& 0.814& 0.912           \\
		\cline{2-12} & DKNet~\cite{DKNet}        & 0.933& 0.973& 0.884& 0.700& \textbf{0.901} & 0.633& \textbf{0.658} & 0.879& 0.945& 0.962\\
		\cline{2-12} & B-Seg (Ours) & \textbf{0.972} & \textbf{0.988} & \textbf{0.967} & \textbf{0.905} & 0.735& \textbf{0.789} & 0.580 & \textbf{0.937} & \textbf{0.976} & \textbf{0.985}  \\
		\midrule
		\multirow{6}{*}{Wuhu} & PointGroup~\cite{jiang2020pointgroup}   & \textbf{0.926} & 0.970& 0.958& 0.928& -& 0.746& 0.655& \textbf{0.959} & 0.950& 0.979\\
		\cline{2-12} & HAIS~\cite{chen2021hierarchical}       & 0.717& 0.898& 0.876& 0.613& -& 0.105& 0.330 & 0.840& 0.813& 0.978\\
		\cline{2-12} & SoftGroup~\cite{vu2022softgroup}   & 0.843& 0.944& 0.929& 0.862& -& 0.479& 0.504 & 0.915& 0.917& 0.935\\
		\cline{2-12} & DyCo3D~\cite{DyCo3D}& 0.811& 0.908& 0.906& 0.648& -& 0.347& 0.419 & 0.869& 0.863& 0.959\\
		\cline{2-12} & DKNet~\cite{DKNet}& 0.919& 0.972& \textbf{0.968} & 0.894& -& 0.675& 0.610 & 0.956& \textbf{0.953} & 0.985\\
		\cline{2-12} & B-Seg (Ours) & 0.882& \textbf{0.973} & 0.955& \textbf{0.934} & -& \textbf{0.785} & \textbf{0.688} & 0.958& 0.930& \textbf{0.990}  \\
		\midrule
		\multirow{6}{*}{Campus} & PointGroup~\cite{jiang2020pointgroup}  & 0.891& 0.939& 0.960& -& \textbf{0.568} & \textbf{0.663} & \textbf{0.819} & 0.895& 0.959& 0.959\\
		\cline{2-12} & HAIS~\cite{chen2021hierarchical}& 0.852& 0.854& 0.885& -& 0.068& 0.231& 0.545 & 0.715& 0.917& 0.904\\
		\cline{2-12} & SoftGroup~\cite{vu2022softgroup}& 0.813& 0.816& 0.841& -& 0.061& 0.134& 0.521 & 0.897& 0.939& 0.949\\
		\cline{2-12} & DyCo3D~\cite{DyCo3D}& 0.631& 0.600& 0.749& -& 0.081& 0.222& 0.476 & 0.436& 0.723& 0.845\\
		\cline{2-12} & DKNet~\cite{DKNet}& 0.942& 0.919& 0.920& -& 0.418& 0.599& 0.805 & 0.890& 0.942& 0.896\\
		\cline{2-12} &
		B-Seg (Ours) & \textbf{0.962} & \textbf{0.957} & \textbf{0.969} & -& 0.498& 0.588& 0.794& \textbf{0.910} & \textbf{0.970} & \textbf{0.968}  \\
		\midrule
		\multirow{6}{*}{Longhua} &
		PointGroup~\cite{jiang2020pointgroup}& 0.868& 0.844& 0.811& 0.745& 0.766& 0.699& 0.696& 0.764& 0.828& 0.374\\
		\cline{2-12} & HAIS~\cite{chen2021hierarchical}& 0.830& 0.870& 0.914& 0.823& 0.090& 0.234& 0.247& 0.529& 0.918& \textbf{0.957}  \\
		\cline{2-12} & SoftGroup~\cite{vu2022softgroup}& 0.697& 0.847& 0.798& 0.768& 0.216& 0.329& 0.416 & 0.846& 0.955& 0.948\\
		\cline{2-12} & DyCo3D~\cite{DyCo3D}& 0.227& 0.815& 0.791& 0.663& 0.142& 0.362& 0.253 & 0.375& 0.798& 0.829\\
		\cline{2-12} & DKNet~\cite{DKNet}& 0.456& 0.847& 0.812& 0.799& 0.872& 0.749& 0.760  & 0.621& 0.831& 0.925\\
		\cline{2-12} & B-Seg (Ours) & \textbf{0.991} & \textbf{0.939} & \textbf{0.984} & \textbf{0.951} & \textbf{0.902} & \textbf{0.789} & \textbf{0.848} & \textbf{0.922} & \textbf{0.979} & 0.924\\
		\bottomrule
	\end{tabular}
 }
	\label{tab:subtable1}
\end{table*}
\begin{table*}[t!]
\caption{3D instance segmentation performance of various methods for different cross-scene training/testing settings.
}
\resizebox{0.9\textwidth}{!}{
\begin{tabular}{c|ccc|ccc|ccc}
\toprule
& \multicolumn{3}{c|}{Train: Qingdao + Wuhu; Test: Longhua} & \multicolumn{3}{c|}{Train: Campus; Test: Qingdao + Wuhu} & \multicolumn{3}{c}{Train: Longhua; Test: Yingrenshi} \\ 
\midrule
Method & 
\makebox[0.08\textwidth][c]{AP}& \makebox[0.08\textwidth][c]{AP50} & \makebox[0.08\textwidth][c]{AP25} & \makebox[0.08\textwidth][c]{AP} & \makebox[0.08\textwidth][c]{AP50} & \makebox[0.08\textwidth][c]{AP25} & \makebox[0.08\textwidth][c]{AP} & \makebox[0.08\textwidth][c]{AP50} & \makebox[0.08\textwidth][c]{AP25} \\ 
\midrule
PointGroup~\cite{jiang2020pointgroup} &0.300&0.482&0.618&0.243&0.374&0.514&0.558&0.660&0.722\\
Hais~\cite{chen2021hierarchical}&0.158&0.265&0.380&0.367&0.493&0.568&0.427&0.530&0.671\\
SoftGroup~\cite{vu2022softgroup}&0.121&0.198&0.310&\textbf{0.391}&0.472&0.540&0.439&0.535&0.566\\
DyCo3D~\cite{DyCo3D}&0.037&0.082&0.356&0.009&0.035&0.244&0.019&0.140&0.411\\
DKNet~\cite{DKNet}&0.139&0.208&0.286&0.075&0.136&0.212&0.297&0.389&0.389\\
B-Seg (Ours) &\textbf{0.323}&\textbf{0.486}&\textbf{0.622}&0.353&\textbf{0.507}&\textbf{0.615}&\textbf{0.621}&\textbf{0.700}&\textbf{0.739}\\ 
\bottomrule
\end{tabular}}
\label{tab:result2}
\end{table*} 

\subsection{Experiment settings}
\vspace*{2mm}
\noindent
\textbf{Implementation Details.}
We use one Quadro RTX 6000 GPU for model training with a batch size of 2 for a single scene and 4 for joint scenes. We use the Adam optimizer~\cite{2014Adam} with an initial learning rate of 0.001 for 400 epochs. For a fair comparison, we follow the original experimental settings of each method and modify only the hyperparameters to suit our dataset,~\eg, the clustering grouping radius for ~\cite{jiang2020pointgroup, chen2021hierarchical, vu2022softgroup}. For the 3D sparse convolution, we set the voxel size as $\frac{1}{3}\times\frac{1}{3}\times\frac{1}{3} m^{3}$. In the training, to balance the GPU memory limit and data block size, we set the maximum number of points as 500,000 and randomly adjust the input size by cropping a block if its size exceeds the maximum, similar to~\cite{jiang2020pointgroup}. At the inference stage, we directly input the whole block into the network.

\vspace*{2mm}
\noindent
\textbf{Evaluations Metrics.}
We employ Average Precision ($AP$, $AP_{25}$, $AP_{50}$) to explore the building instance segmentation. $AP_{25}$ and $AP_{50}$ are the $AP$ scores with an Intersection-over-Union (IoU) threshold of 25$\%$ and 50$\%$, respectively. $AP$ is the mean Average Precision score from the IoU threshold of 25$\%$ to 95$\%$ with a step of 5$\%$. Besides, we use mean Intersection-over-Union (mIoU) to evaluate the instance segmentation performance on the fine-grained building categories.

\subsection{Experimental results}
We evaluate our UrbanBIS dataset and the B-Seg method using some mainstream and representative 3D instance segmentation methods. 
From Sec.~\ref{sec:comparison}, we can see the large variations across different scenes in the data, so we set up different training settings to explore the method's effectiveness for different situations. 

\begin{figure*}[p]
	\centering
	\includegraphics[width=0.94\linewidth]{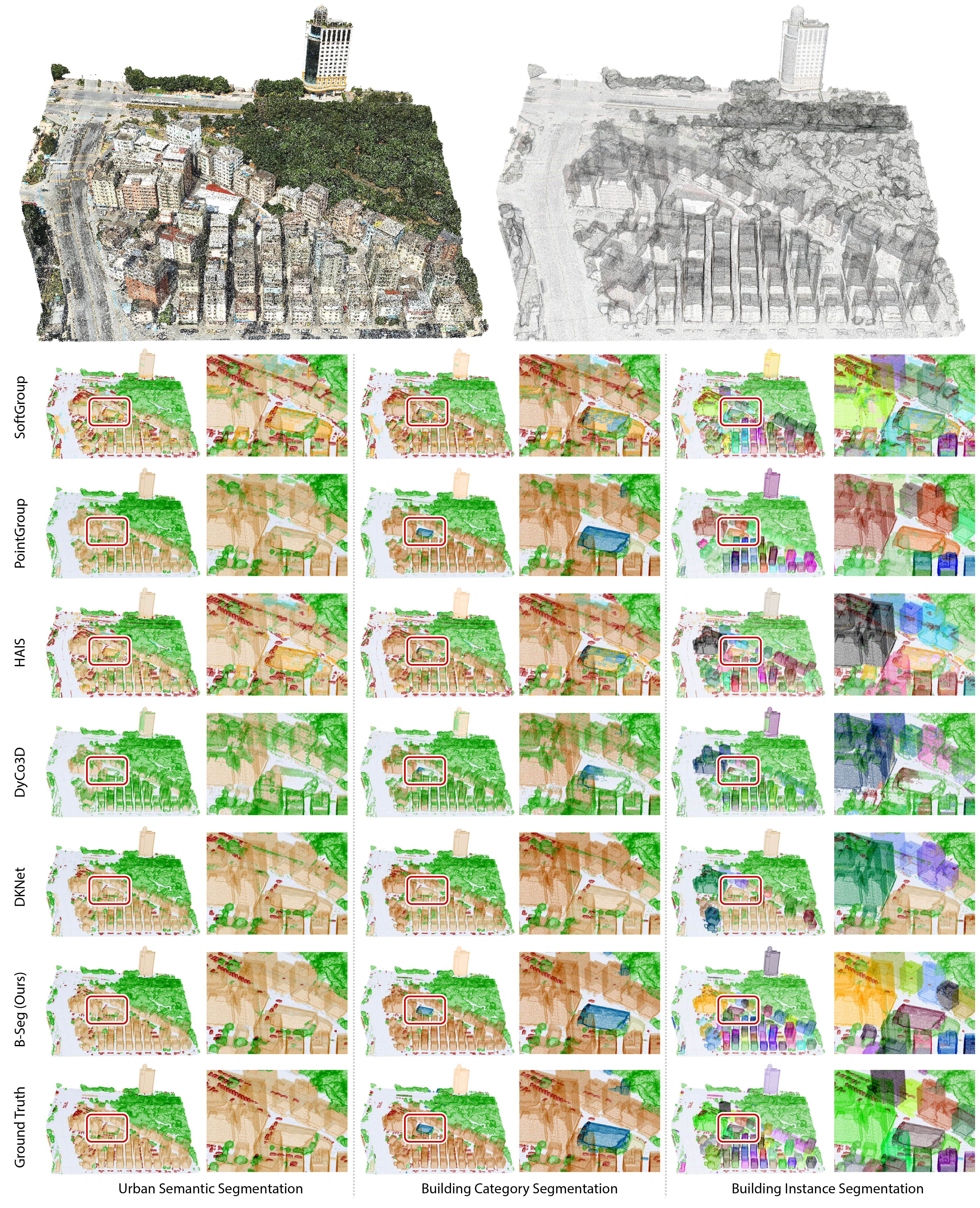}
	\caption{Qualitative results of B-Seg and comparison methods on a test block in the Longhua scene. From left to right is the urban semantic segmentation, building category segmentation, and building instance segmentation. The prediction of B-Seg shows more accurate segmentation results in this dense scene.}
	\label{fig: result_longhua}
\end{figure*}

\vspace*{2mm}
\noindent
\textbf{Building Instance Segmentation Performance.}
First, both the train and test sets come from the same scene.
As reported in Table~\ref{tab:result1},
we can see that B-Seg achieves the best performance over the SOTA 3D point cloud segmentation methods \emph{consistently for all the metrics in all the scenes}.
Also, B-Seg has a faster processing speed than others; see the T columns.
The other methods~\cite{jiang2020pointgroup, chen2021hierarchical, DyCo3D, vu2022softgroup} are clustering-based, thus requiring a time-consuming point-wise grouping process.
For the very recent method DKNet~\cite{DKNet}, though it is clustering-free, it needs to encode every instance proposal into a kernel feature for the dynamic convolution.
While DyCo3D~\cite{DyCo3D} also adopts a dynamic-convolution-based model, it still requires a point-wise grouping first. 
On the contrary, B-Seg adopts a simple but effective strategy by aggregating points with a relation matrix and merging points only through a small set of candidates.
The visualization result can be seen in Fig. \ref{fig: result_longhua}.

\vspace*{2mm}
\noindent
\textbf{Fine-grained Building Instance Segmentation Performance.}
We further evaluate the segmentation performance of all methods on the building sub-categories. As Table~\ref{tab:subtable1} shows, B-Seg performs favorably over other methods for most sub-categories, except that it performs slightly worse than PointGroup~\cite{jiang2020pointgroup} on three categories in the Campus.
These categories are overall harder to distinguish and their features are more similar in Campus, so requiring a stronger feature extractor. Note also that we do not have results for some categories in Wuhu and Longhua (see ``-'' in the table), as the number of associated buildings in these cases is either too small or not existent.
From the experiments, we can see the strength of B-Seg in analyzing buildings in 3D urban scenes and its potential for supporting building-aware 3D machine learning tasks.

\vspace*{2mm}
\noindent
\textbf{Joint-training on all scenes.}
We train the methods collectively on samples of all scenes. As Table~\ref{tab:joint-training} shows, B-Seg still achieves satisfactory results in average performance with the fastest speed.

\vspace*{2mm}
\noindent
\textbf{Cross-scene Building Instance Segmentation Performance.}
Next, we explore the model generalization capability of various methods by training each method on some scenes and testing its trained model on others. As Table~\ref{tab:result2} shows, B-Seg is able to achieve better cross-scene 3D instance segmentation performance than the SOTA methods for the different combinations of train and test scenes.
Overall, the experiment demonstrates the strong generalization capability of B-Seg. Such a capability is particularly important for handling \emph{unseen} urban scenes.

\vspace*{2mm}
\noindent
\textbf{Test on InstanceBuilding.}
To further explore B-Seg, we conducted experiments on InstanBuilding~\cite{chen2022mvs}, a triangle mesh dataset with building instance annotations primarily for segmenting adjacent buildings. Table~\ref{tab:newdataset} reports the results, in which the results of~\cite{chen2022mvs} are directly copied from original paper of InstanceBuilding. In terms of AP, B-Seg does not show obvious disadvantages and even outperforms other methods on Scene 2. Yet, B-Seg does not use any relevant image data or depth information at all, showing that B-Seg does have advantages in dealing with unknown 3D scenes and can achieve good performance.

It can be seen that B-Seg has advantages in dealing with the instance segmentation in urban scenes from the above results, including faster processing speed, better segmentation performance and more robust generalization. This is because B-Seg does not rely on the training set for the clustering makes it less prone to over-fitting and ensures its generalizability in different urban scenes.

\begin{table}[t!]
\caption{Joint training on all scenes in UrbanBIS. We combine all the training and validation samples to train all the methods below, and employ all the test samples in the evaluation. The performance for fine-grained buildings can be found in supplementary material.}
\vspace{-0.3cm}
\resizebox{0.4\textwidth}{!}{
\begin{tabular}{c|ccc|c}
\toprule
Method & AP & AP50 & AP25 & Time (s) \\ \midrule
PointGroup~\cite{jiang2020pointgroup} & 0.377 & 0.549 & 0.664 & 6.918 \\
HAIS~\cite{chen2021hierarchical} & 0.373 & 0.515 & 0.587 & 4.500 \\
SoftGroup~\cite{vu2022softgroup} & \textbf{0.402} & 0.490 & 0.560 & 4.389\\
DyCo3D~\cite{DyCo3D} & 0.129 & 0.246 & 0.487 & 2.386 \\
DKNet~\cite{DKNet} & 0.271    & 0.348    & 0.408  & 1.059 \\
B-Seg (Ours) & 0.401 & \textbf{0.551} & \textbf{0.671} & \textbf{0.969} \\ \bottomrule
\end{tabular}}
\label{tab:joint-training}
\vspace{-0.1cm}
\end{table}

\begin{table}[t!]
\caption{Experimental results on the InstanceBuilding dataset.
Note that B-Seg does not need images and uses only 3D data without color.}
\vspace{-0.3cm}
\resizebox{0.45\textwidth}{!}{
\begin{tabular}{c|ccc|ccc|ccc}
\toprule
\multirow{2}{*}{Scene} & \multicolumn{3}{c|}{\cite{chen2022mvs} (RBG)} & \multicolumn{3}{c|}{\cite{chen2022mvs} (RGBH)} & \multicolumn{3}{c}{B-Seg (Ours)} \\ \cline{2-10} & AP & AP50 & AP75 & AP & AP50 & AP75 & AP & AP50 & AP75 \\ \midrule 
Scene1  & 0.648 & 0.832 & 0.665  & 0.713  & 0.891 & 0.773 & 0.561 & 0.725 & 0.594    \\
Scene2 & 0.571 & 0.748 & 0.597 & 0.665 & 0.840 & 0.681 & 0.764 & 0.930 & 0.782    \\
Scene3  & 0.636 & 0.888 & 0.693 & 0.667 & 0.929 & 0.717 & 0.599 & 0.820 & 0.625    \\
Scene4 & 0.653 & 0.857  & 0.699 & 0.673 & 0.876 & 0.711 & 0.519 & 0.739 & 0.491    \\ \bottomrule
\end{tabular}}
\vspace{-0.3cm}
\label{tab:newdataset}
\end{table}

\section{Discussion and Conclusion}

In this paper, we presented UrbanBIS, a large-scale 3D urban dataset, providing six real-world urban scenes of 2.5 billion annotated point samples over a vast area of $10.78km^2$.
Beyond the existing datasets, it is a large-scale 3D real-world urban dataset, which is multi-functional with multiple data formats: dense semantic annotations on the point clouds and meshes, fine-grained instance and semantic segmentation on the buildings, and high-quality 3D reconstruction models, as well as high-resolution aerial-acquisition photos.
UrbanBIS comprises scenes of different urban styles and building compositions, so different portions of the dataset can be employed for different purposes of study.
Further, we develop B-Seg, an end-to-end framework to segment the large point clouds in UrbanBIS.
Compared with mainstream methods, B-Seg demonstrates better accuracy and faster inference speed on UrbanBIS. In the future, we shall release UrbanBIS and B-Seg.
\begin{acks}
We thank the anonymous reviewers for their constructive comments. This work was supported in parts by NSFC (U21B2023, 62161146005, U2001206), GD Talent Program (2019JC05X328), DEGP Innovation Team (2022KCXTD025), Shenzhen Science and Technology Program (KQTD20210811090044003, RCJC20200714114435012, JCYJ20210324120213036), and Guangdong Laboratory of Artificial Intelligence and Digital Economy (SZ).
\end{acks}

\bibliographystyle{ACM-Reference-Format}
\bibliography{UrbanBIS_ref}

\end{document}